\documentclass[aps, prb, superscriptaddress, twocolumn, amsfonts, amsmath, amssymb, floatfix]{revtex4-2}
\usepackage{graphicx}
\usepackage{float}
\usepackage{sidecap}
\usepackage{dcolumn}
\usepackage{bm}
\usepackage{epsfig}
\usepackage{braket}
\usepackage{color}
\usepackage{ulem}
\usepackage{amsmath}
\usepackage{caption}
\usepackage{subcaption}
\usepackage[toc,page]{appendix}
\usepackage[colorlinks=true, letterpaper=true, pdfstartview=FitV, linkcolor=blue, citecolor=blue, urlcolor=blue]{hyperref}

\begin{document}
\title{Quantum transport in a multi-path graphene Aharonov-Bohm inteferometer}

\author{Cynthia I. Osuala}
\affiliation{Department of Physics, Stevens Institute of Technology, Hoboken, NJ 07030, USA}

\author{Zitao Tang}
\affiliation{Department of Mechanical Engineering, Stevens Institute of Technology, Hoboken, NJ 07030, USA}

\author{Stefan Strauf}
\affiliation{Department of Physics, Stevens Institute of Technology, Hoboken, NJ 07030, USA}
\affiliation{Center for Quantum Science and Engineering, Stevens Institute of Technology, Hoboken, NJ 07030, USA}

\author{Eui-Hyeok Yang}
\affiliation{Department of Mechanical Engineering, Stevens Institute of Technology, Hoboken, NJ 07030, USA}
\affiliation{Center for Quantum Science and Engineering, Stevens Institute of Technology, Hoboken, NJ 07030, USA}

\author{Chunlei Qu}
\email{Corresponding author. cqu5@stevens.edu}
\affiliation{Department of Physics, Stevens Institute of Technology, Hoboken, NJ 07030, USA}
\affiliation{Center for Quantum Science and Engineering, Stevens Institute of Technology, Hoboken, NJ 07030, USA}

\date{\today}
\begin{abstract}
We investigate the quantum transport dynamics of electrons in a multi-path Aharonov-Bohm interferometer comprising several parallel graphene nanoribbons. At low magnetic field strengths, the conductance displays a complex oscillatory behavior stemming from the interference of electron wave functions from different paths, reminiscent of the diffraction grating in optics. With increasing magnetic field strength, certain nanoribbons experience transport blockade, leading to conventional Aharonov-Bohm oscillations arising from two-path interference. We also discuss the impact of edge effects and the influence of finite temperature. Our findings offer valuable insights for experimental investigations of quantum transport in multi-path devices and their potential application for interferometry and quantum sensing.
\end{abstract}

\maketitle

\section{Introduction}
Graphene is a compelling contender for next-generation compact and high-performance electronic devices due to its exceptional properties, including the atomically thin monolayer structure, linear dispersion relation near the Dirac points, absence of energy gap, and high carrier mobility~\cite{geim2007rise, RevModPhys.81.109}. These properties have enabled a myriad of innovative applications ranging from ultrafast electronics and flexible optoelectronics to advanced sensing technologies. Recent research has also unveiled graphene's potential as a versatile playground for exploring intriguing quantum interference phenomena similar to those observed in optics, yet operating on the nanoscale with matter waves. Notable examples of electron optics with graphene include the realization of Fabry-P\'{e}rot interferometers, Mach-Zehnder interferometers, and Veselago lens, which have paved the way for fabricating chip-scale electronic interferometers~\cite{Nat.Nanotechno.2021, doi:10.1126/sciadv.1700600, DOI:10.1126/science.1138020}.

The Aharonov-Bohm (AB) effect is a quintessential example of quantum interference that has garnered substantial attention across various materials systems, including metals~\cite{PhysRevLett.54.2696}, semiconductor heterostructures~\cite{PhysRevLett.55.2344}, carbon nanotubes~\cite{Bachtold}, and topological insulators~\cite{peng2010, PhysRevLett.124.126804, Huang2020, behner2023aharonov}. 
It arises from the interference of wave functions of charged particles encircling a ring structure in the presence of a perpendicular magnetic field. Due to the presence of magnetic flux, charged particles pick up a different phase $\Delta\varphi = 2\pi BS/\phi_0$ when traversing along the two paths where $\phi_0=h/e$ denotes the flux quantum, $S$ is the area enclosed by the ring, $B$ is the strength of the applied perpendicular magnetic field~\cite{PhysRev.115.485} . Consequently, by varying the magnetic field, the conductance exhibits a periodic oscillation with an oscillation frequency given by $f_B=S/\phi_0$. AB oscillations have been observed in various graphene materials, including mechanically exfoliated graphene~\cite{https://doi.org/10.1002/pssb.200982284,  PhysRevB.77.085413, 10.1063/1.4717622, dauber2017aharonov, Huefner_2010}, epitaxial graphene~\cite{SCHELTER20121411}, CVD graphene~\cite{ Nguyen2019AharonovBohmII, 10231188},  exfoliated bilayer graphene~\cite{PhysRevB.78.045404, PhysRevB.81.045431}. However, most investigations have focused on two-path interferometry setups, which inherently limit the complexity of achievable interference patterns~\cite{Wurm_2010,PhysRevB.94.195315,PhysRevB.88.035408,PhysRevB.95.125418}. The natural question is, what happens when considering a multi-path interferometer involving
interconnected quantum pathways? Despite its conceptual importance and simplicity, multi-path AB oscillations have been relatively underexplored in the literature. 

Here, we systematically investigate quantum transport in a multi-path AB interferometer comprising several parallel graphene nanoribbons (GNRs). We found that the magnetoconductance exhibits complex oscillatory behavior due to the intricate interplay between electron trajectories, magnetic flux, and the quantum Hall effect. When the magnetic field is weak, electrons can be transported from source to drain through all pathways, giving rise to AB interference reminiscent of the diffraction grating effect in optics. As the magnetic field increases, the bulk current evolves into chiral edge states due to the quantum Hall effect, resulting in unidirectional flow and the obstruction of specific pathways. In this regime, we demonstrate a dramatic change in the AB oscillation that evolves from multi-path to two-path interference. Our results showcase the complexity achievable in multi-path interferometers and offer new avenues for harnessing the potential of graphene-based systems for quantum-enhanced technologies.

\begin{figure*}
\centering
\includegraphics[width=0.8\textwidth]{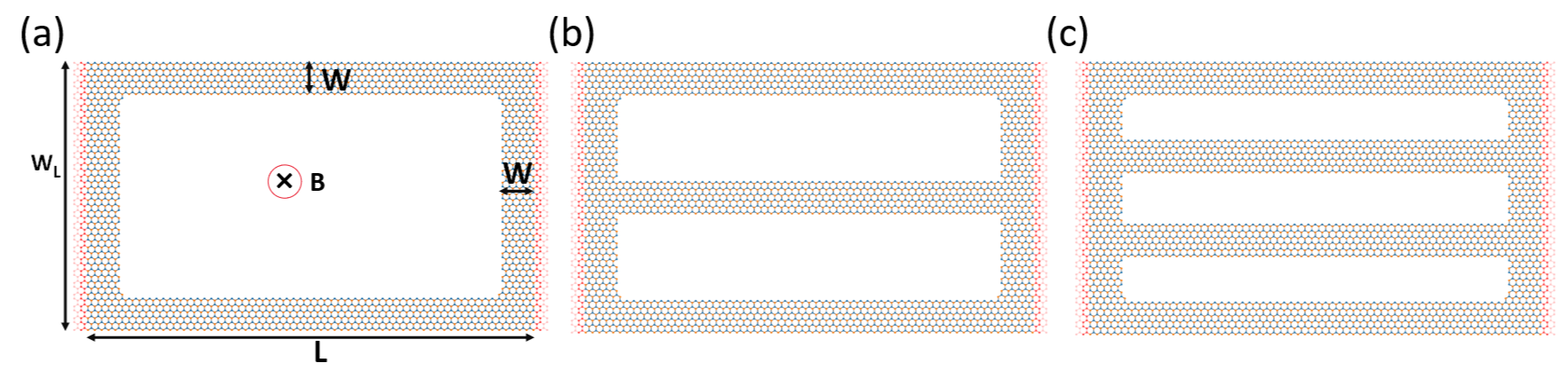}
\captionsetup{singlelinecheck=false, justification=justified}
\caption{\justifying Schematic illustration of our GNR rings subjected to a uniform magnetic field along the perpendicular direction with (a) two paths, (b) three paths, and (c) four paths. The length is denoted as $L=500$nm, the width of the lead is $W_L=300$nm, and the width of the ring arm is $W=40$nm. The ring arms have the same width and enclose the same total area.}
\label{fig:sys}
\end{figure*}

\section{Quantum Transport at $T=0$}
The carbon atoms of graphene arrange themselves in a 2D honeycomb lattice, with two atoms per unit cell, giving rise to remarkable properties and behaviors of graphene. To compute the electronic transport in GNR rings we employ the following spinless nearest-neighbor tight-binding model~\cite{RevModPhys.81.109}
\begin{equation} H=\sum_{i}\epsilon_{i}c_{i}^{\dag}c_{i} + \sum_{\langle i,j \rangle}t_{ij}(c_{i}^{\dag}c_{j} + h.c.)
\label{eq:1}
\end{equation}
where $c_{i}^{\dag}$ ($c_{i}$) represents the creation (annihilation) operator for an electron at the $i$-th lattice site, with an associated on-site potential energy $\epsilon_{i}$ which we have set to zero in this work.
The summation $\sum_{\langle i,j \rangle}$ is limited to the nearest-neighbor atoms on the honeycomb lattice with a lattice constant of $a_0 = 0.246$ nm. 
Under the influence of a uniform perpendicular magnetic field $\mathbf {B} = (0, 0, B)$,  the tight-binding Hamiltonian $H$  is modified through the Peierls substitution~\cite{Peierls1933} 
\begin{equation} t_{ij} = t_{0} e^{-i\frac{2\pi}{\phi_{0}} \int_{\mathbf{r}_{i}}^{\mathbf{r}_{j}} \mathbf{A}(\mathbf{r}) \cdot d\mathbf{r}}
\end{equation}
where $t_0 \approx 2.7$ eV is the hopping parameter and $\mathbf{A}$ is the gauge field specifically in the Landau gauge as $\textbf{A} = (-By, 0, 0)$. To enhance computational efficiency, we employ a scaling factor of $s = 5$ with the hopping parameter rescaled to $t = t_0/s$ and lattice spacing $a = a_{0}s$~\cite{PhysRevLett.114.036601}. At zero temperature the conductance is calculated using the Landauer formula $G = (e^2/h)\sum_{\alpha \beta}\mid t_{\alpha \beta} \mid^2$ with $t_{\alpha\beta}$ representing the probability amplitude for the transmission from mode $\beta$ in the input lead to mode $\alpha$ in the output lead. In this work, we employ the Kwant package to numerically simulate the quantum transport dynamics of the multi-path graphene system~\cite{Groth_2014}.

Figure ~\ref{fig:sys} depicts a parallel zigzag GNR ring with multiple paths for electron matter wave interference. Figure~\ref{fig:sys}(a) displays the configuration with two paths, which we will refer to as a 2slits system. Furthermore, we introduce a third and fourth pathway, yielding a 3slits system (Fig.~\ref{fig:sys}(b)) and a 4slits system (Fig.~\ref{fig:sys}(c)). Two semi-infinite leads, indicated in red, are symmetrically attached to each end of the ribbon. The dimensions of the rectangular zigzag GNR are defined by the parameters displayed in Fig.~\ref{fig:sys}(a). 

\begin{figure}[b!]
    \centering    \includegraphics[width=0.5\textwidth]{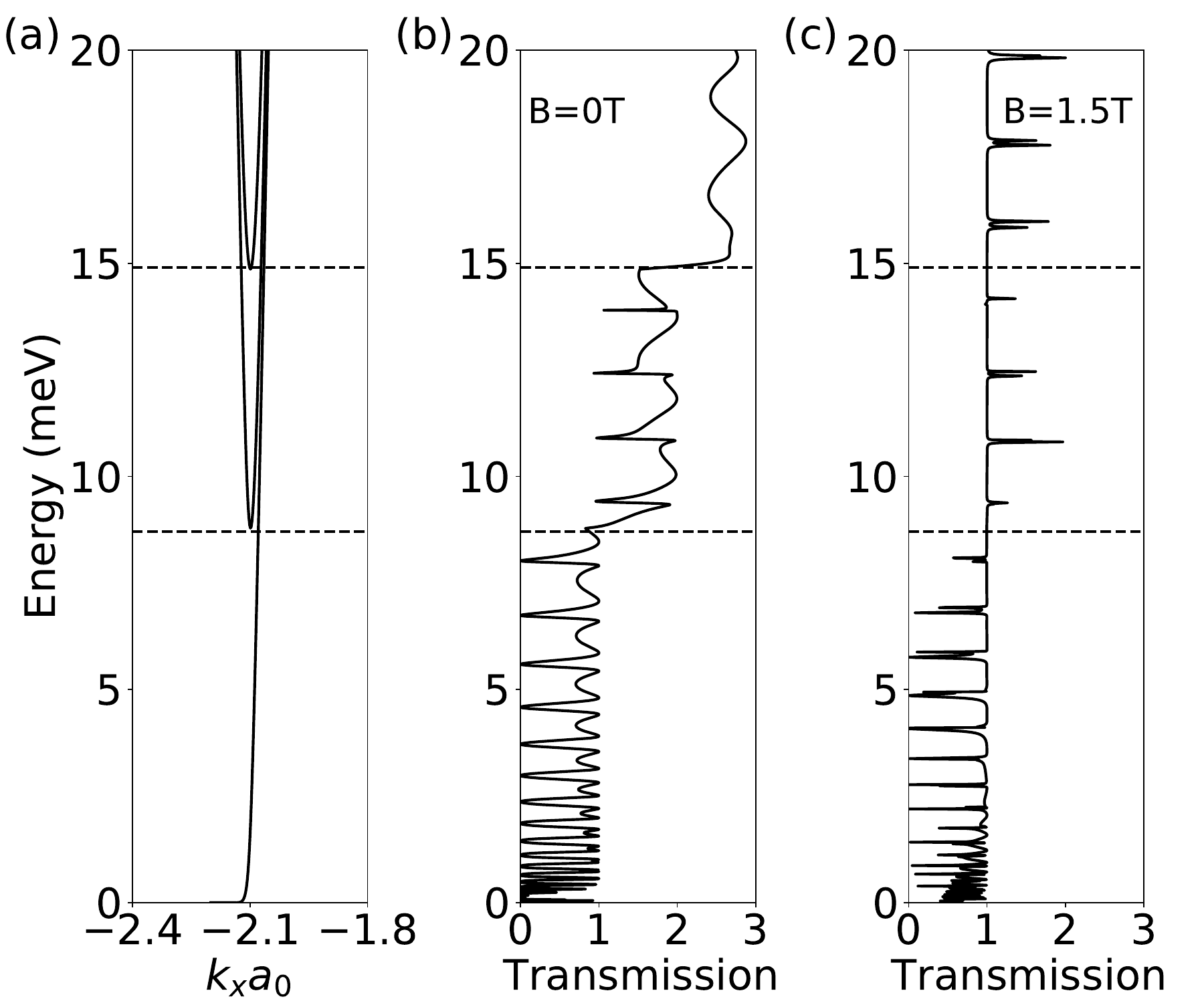}   \captionsetup{singlelinecheck=false, justification=justified}
    \caption{\justifying (a) Band structure of the lead and (b, c) the corresponding transmission probability of the 3slits system at $B=0$T and $B=1.5$T, respectively
   }.
\label{fig:transm}
\end{figure}

In Fig.~\ref{fig:transm}(a), we present the band structure of the lead (which is the same for the three configurations) with a zigzag edge at $B=0$T. Only the lowest three conduction subbands are shown as they are relevant to the quantum transport dynamics. The corresponding transmission probability for the 3slits system is depicted in Fig.~\ref{fig:transm}(b). Within the low-energy regime, corresponding to the first subband, a distinct transmission probability pattern, characterized by periodic oscillations of varying amplitudes, emerges. This pattern diminishes as the Fermi energy is increased to cross the second subband. Upon further increase in the Fermi energy, the transmission probability demonstrates regular oscillations similar to those observed in the 2slits system. An interesting phenomenon happens at B=1.5 T. In this scenario, the oscillatory pattern exhibited at low Fermi energies undergoes a process of smearing out and suppression as the Fermi energy increases. At higher Fermi energies, multiple sub-bands contribute to the transport dynamics of electrons, leading to interference patterns becoming less pronounced or even smeared out. In the following, we shall focus on the low Fermi energy regime.

\begin{figure}[t!]
    \centering
    \includegraphics[width=0.5\textwidth]{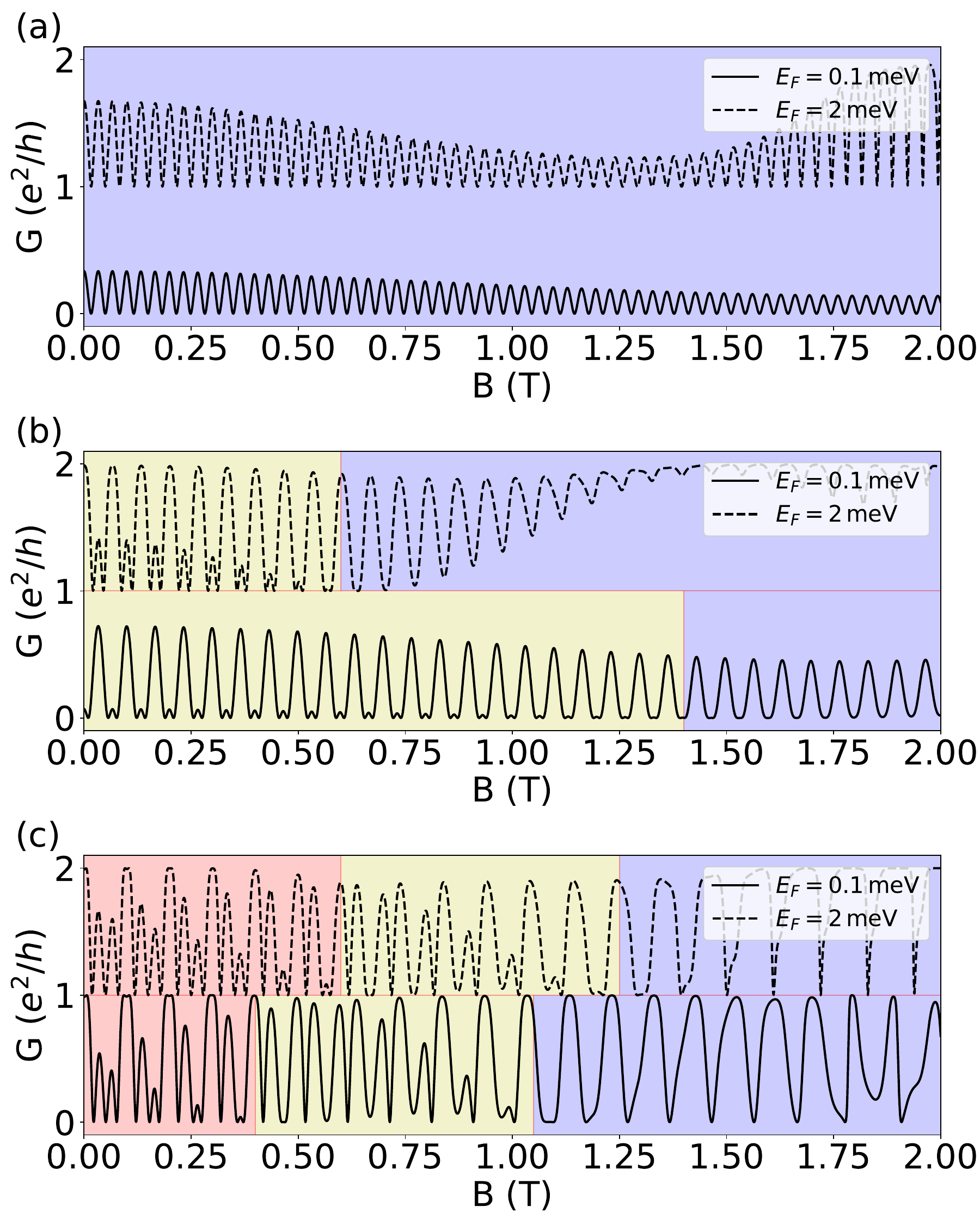}
    \captionsetup{singlelinecheck=false, justification=justified}
    \caption{\justifying  
    Conductance $G$ as a function of the magnetic field $B$ for Fermi energies $E_F = 0.1$meV (solid line) and $E_F = 2$meV (dashed line) at temperature $T = 0$ for the GNR rings with (a) 2 slits, (b) 3 slits, and (c) 4 slits. The purple-shaded area corresponds to interference between two paths, the yellow-shaded area corresponds to interference between three paths, and the red-shaded area corresponds to interference between four paths. The curves for $E_F = 2$meV have been shifted by $1e^2/h$ for improved visualization.}
    \label{fig:cond}
\end{figure}

In Fig.~\ref{fig:cond}, the conductance $G$ is presented as a function of the magnetic field $B$ for nanoribbon rings. The conductance plot in Fig.~\ref{fig:cond}(a) for the 2slits system exhibits regular oscillations that arise from the two-path AB effect. The oscillation period $\Delta B$ is approximately $34.5$mT, in excellent agreement with the theoretical expression $\phi_0/S$ where $S\approx L\times W_L$. The conductance plots in Fig.~\ref{fig:cond}(b, c) exhibit notable and distinctive features. In the case of the 3slits configuration, two types of oscillation patterns are observed. Within the yellow-shaded region, the amplitudes of the conductance oscillation exhibit an alternating trend, where they undergo systematic increments and decrements as a function of the magnetic field strength. This behavior is prominent for very small values of B. However, as the magnetic field strength increases beyond a certain threshold, which depends on the Fermi energy $E_F$ of the incident electron, a regular oscillation pattern emerges (indicated by the purple-shaded region), corresponding to the behavior observed in the 2slits system. For the 4-slits system, three types of oscillation patterns can be identified, as indicated by the three different colors in Fig.~\ref{fig:cond}(c). The conductance clearly consists of three oscillation frequencies at a very low magnetic field. As the magnetic field strength is increased, the behavior evolves into that for the 3slits and 2slits systems, respectively.

\begin{figure}[t!]
    \centering
    \includegraphics[width=0.5\textwidth]{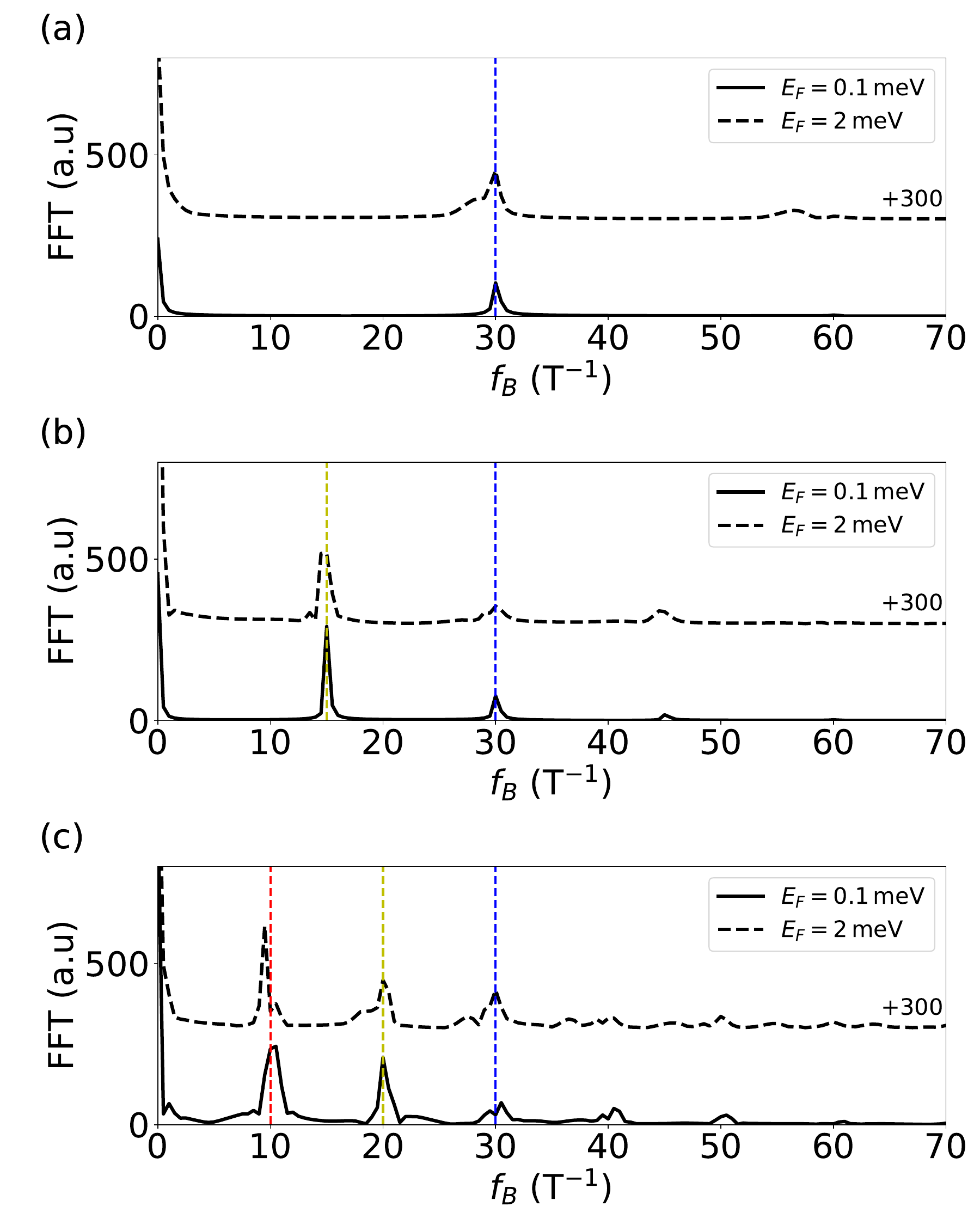}
    \captionsetup{singlelinecheck=false, justification=justified}
    \caption{\justifying 
    Fourier transform of the magnetoconductance oscillations shown in Fig.~\ref{fig:cond}, with blue, yellow, and red vertical lines indicating the frequencies resulting from interference between the top and bottom slits,  neighboring slit (b) and next nearest-neighboring slits (c), and neighboring slits, respectively.}
\label{fig:FFT}
\end{figure}

To quantitatively analyze the complex magnetoconductance oscillations presented in Fig.~\ref{fig:cond}, we have performed a fast Fourier transform (FFT). As shown in Fig.~\ref{fig:FFT}(a), the conductance oscillations for the 2slits system exhibit a prominent peak at a frequency of 30/T, which agrees with the conventional AB oscillations arising from electron matter wave interference along the two pathways. Higher harmonics, such as the peak at around 60/T, are also visible. For the 3slits system, in addition to the 30/T frequency component, other frequencies emerge, with the one at 15/T dominating. This suggests that the magnetoconductance oscillation of the 3slits system (Fig.~\ref{fig:cond}(b)) results from two types of interference: one between neighboring slits, resulting in slower oscillations at 15/T, and the other between the top and bottom slits, yielding faster oscillations at 30/T. The latter frequency component diminishes at relatively higher magnetic field strengths (indicated by the purple-shaded area in Fig.~\ref{fig:cond}(b)), implying that either the top or bottom slit becomes insulating. Similarly, in Fig.~\ref{fig:cond}(c), three distinct frequency components emerge at (i) 10/T, (ii) 20/T, and (iii) 30/T, correspondingly to interference between electron matter waves (i) among neighboring slits, (ii) among the next nearest-neighboring slits, and (iii) between the top and bottom slits, respectively. The evolving oscillations in Fig.~\ref{fig:cond}(c) indicate that one, and subsequently two, pathways of the 4slits system become insulating under intermediate and strong magnetic field strengths.

To gain further insight into the conductance behavior in Fig.~\ref{fig:cond}, we examine the current flow of the whole system under various magnetic field strengths. The current between two sites is defined as~\cite{PhysRevB.94.195315, Gupta2023FractionalFP}
 \begin{equation}
     J_{ij}= i\sum_{\alpha} (\psi_{\alpha i}^{\ast}t_{ij}\psi_{\alpha j} - \psi_{\alpha j}^{\ast}t_{ij}\psi_{\alpha i})
 \end{equation}
where $\psi_{\alpha i}$ is the wave function at the $i$-th site, and $\alpha$ denotes the index of the conducting channels of the two leads, spanning all available energy modes up to the Fermi energy $E_F$.

\begin{figure}[t!]
\centering
\includegraphics[width=0.5\textwidth]{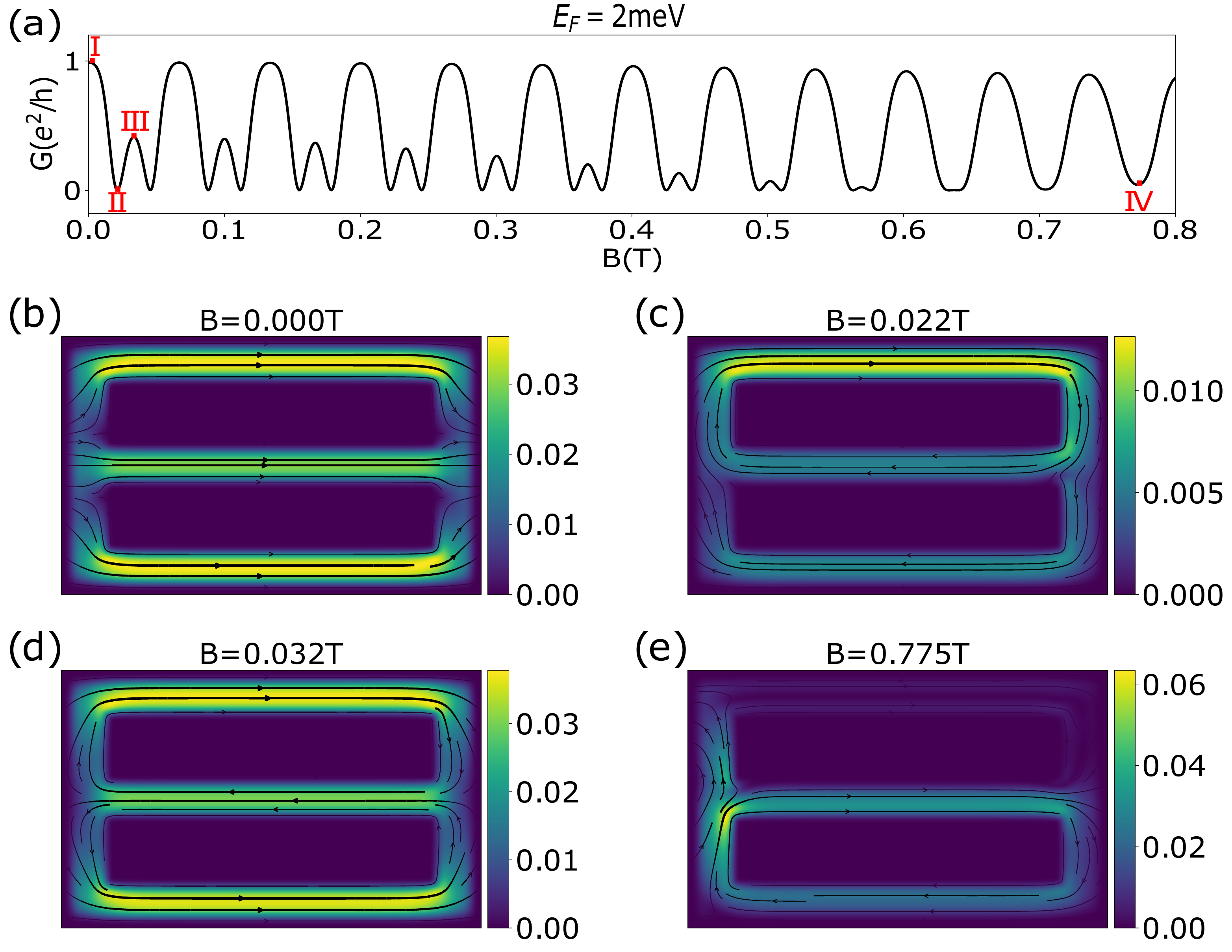}
\captionsetup{singlelinecheck=false, justification=justified}
\caption{\justifying 
(a)  Conductance G as a function of the magnetic field B for the 3slit configuration at $E_F=2meV$. The corresponding current distributions are illustrated in (b-e), each associated with distinct magnetic field strength. Electrons enter the ribbon from the left lead. Darker blue shades depict lower current density with narrower streamlines, while brighter yellow hues represent higher current density with wider streamlines. This interplay of streamlines and colors visually illustrates flow speed and current distribution within the system.}
\label{fig:current}
\end{figure}

Figure~\ref{fig:current} presents the current flow distribution at four different magnetic field strengths, corresponding to the four points denoted as I, II, III, and IV of the 3slits magnetoconductance oscillation curve(Fig.~\ref{fig:current}(a)), respectively. At $B=0$T, the currents in the three pathways flow in the same direction, giving rise to the maximum flow of current and the highest conductance amplitude (Fig.~\ref{fig:current}(b)). In analogy to the optical grating effect, this scenario corresponds to the constructive interference of electron matter waves along the three paths. At $B=0.022$T, the current flows to the right lead via the top pathway and returns back through the two lower pathways, leading to a minimal net current and vanishing conductance(Fig.~\ref{fig:current}(c)). This corresponds to the destructive interference of the electron matter waves along the three paths. At $B=0.032$T, as the Peierls phase continues to accumulate, the current in the lowest path is reversed again, leading to partial constructive interference among the three paths and finite conductance(Fig.~\ref{fig:current}(d)). As the magnetic field increases, the grating effect gradually fades out, and the current along the top pathway is blocked(Fig.~\ref{fig:current}(e)). This is related to the formation of Landau levels and the associated chiral edge state that flows along the lower part of the GNR. When the magnetic field is increased to even larger values, the middle slit will also be blocked due to the further localization of the edge state.

\begin{figure}[t!]
\centering
\includegraphics[width=0.5\textwidth]{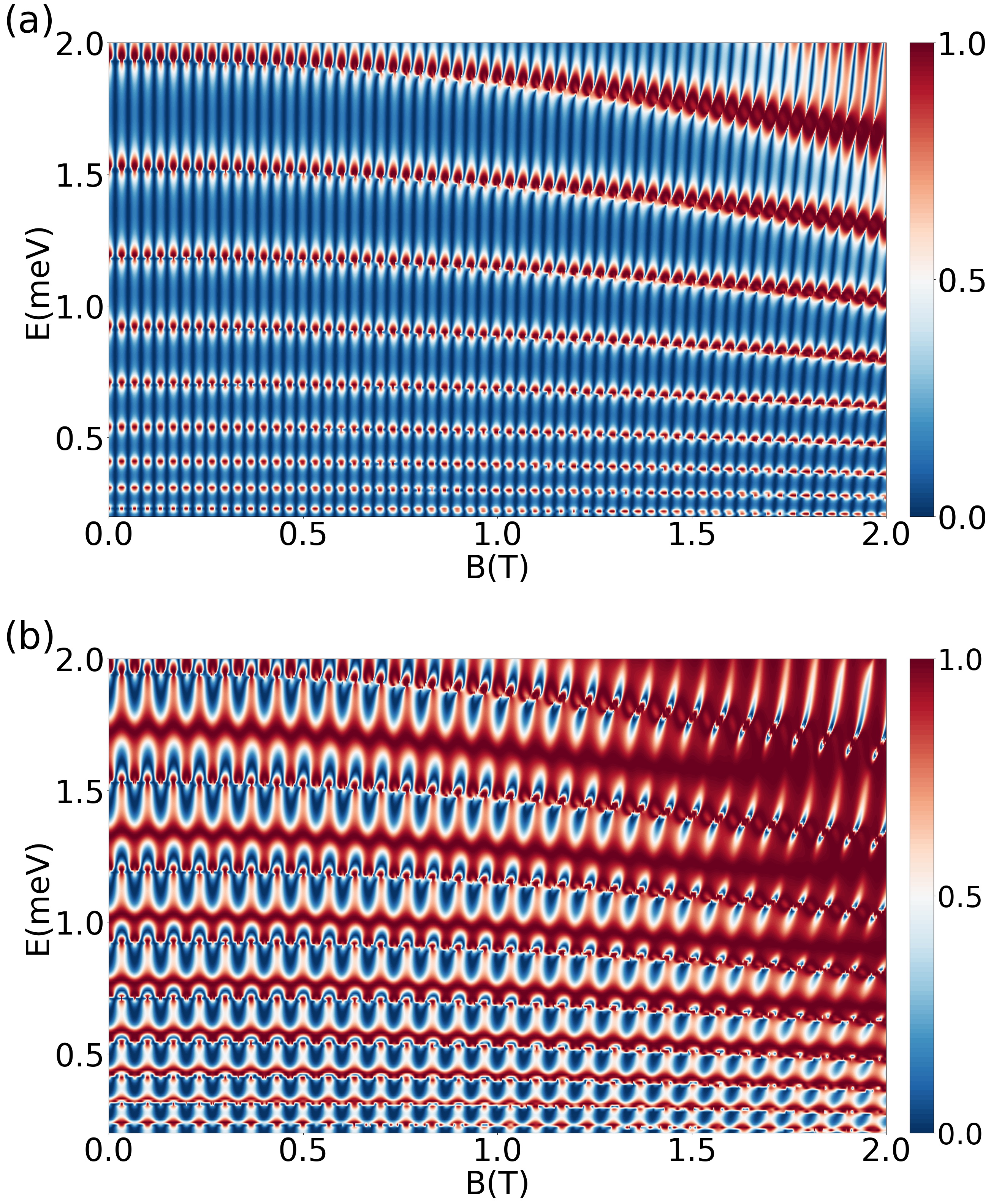}
\captionsetup{singlelinecheck=false, justification=justified}
\caption{\justifying Contour plot of the conductance (in unit of $e^2/h$) as a function of the Fermi energy $E$ and the magnetic field strength $B$ for the (a) 2slits and (b) 3slits configurations.}
\label{fig:trans3D}
\end{figure}

Figure~\ref{fig:trans3D} shows a contour plot of the conductance as a function of Fermi energy and magnetic field strength. In the case of the 2slits system, the conductance displays pronounced AB oscillation, with a frequency that is largely unaffected by changes in Fermi energy. This phenomenon arises from our choice of Fermi energies near the Dirac point. When we vary the Fermi energy while keeping the magnetic field constant, the conductance of the 2slits system exhibits additional oscillations. These oscillations are attributed to the resonant tunneling through the bound states within the GNR, a consequence of confinement along the longitudinal direction. In contrast, the 3slits system exhibits a more complex oscillatory pattern owing to the multi-path interference effects. A diffraction grating effect is evident across a wide spectrum of Fermi energies.

\begin{figure}[t!]
\centering    \includegraphics[width=0.5\textwidth]{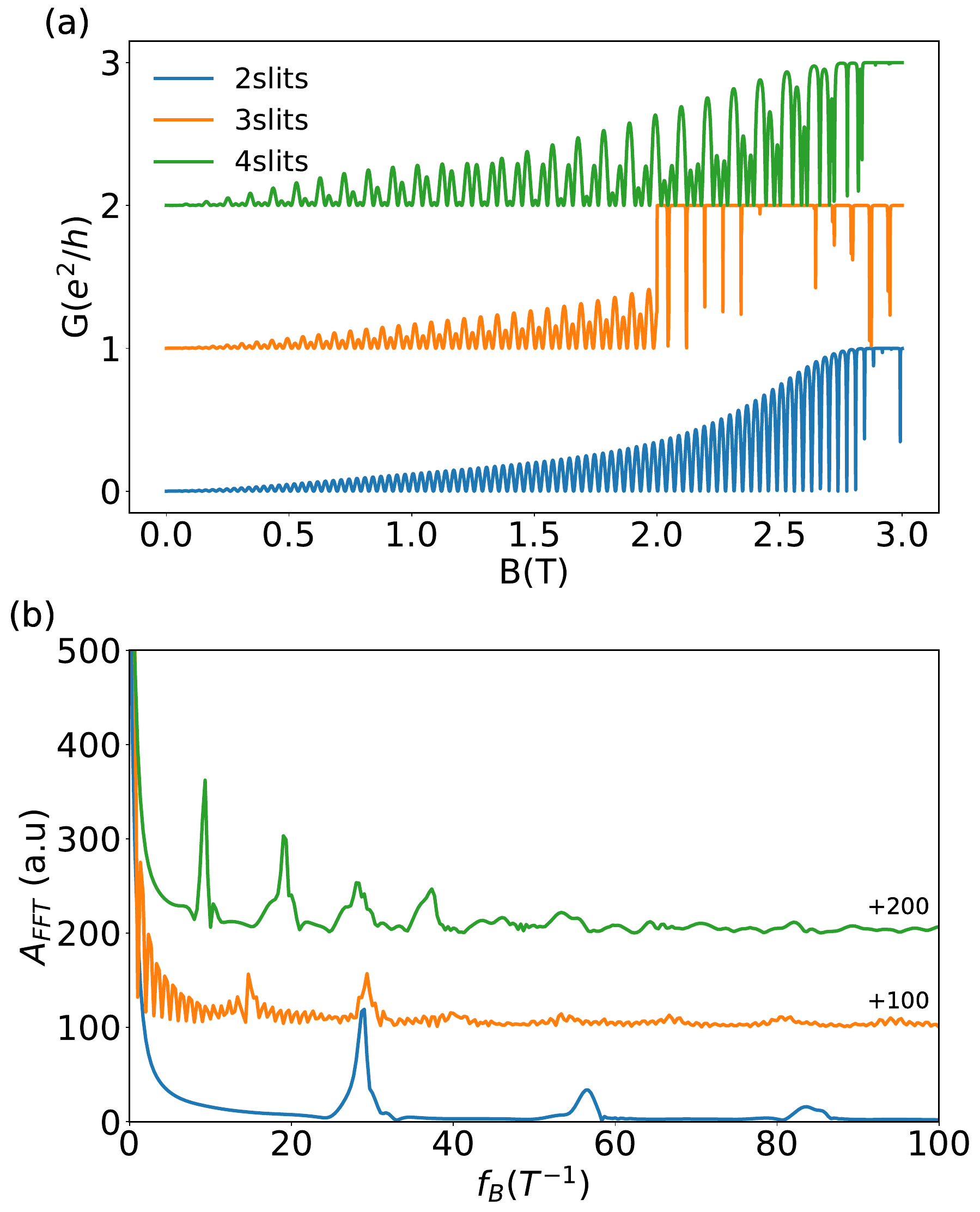}
\captionsetup{singlelinecheck=false, justification=justified}
\caption{\justifying 
(a) Conductance as a function of the magnetic field for armchair GNR rings in the setup depicted in Fig.~\ref{fig:sys} and (b) the corresponding Fourier spectrum for Fermi energies $E_F=2.5$meV. The curves in (a) have been shifted by $1e^2/h$ for improved visualization.}
\label{fig:armchair}
\end{figure}

In our previous analysis, we explored the GNR with a zigzag edge, as depicted in Fig.~\ref{fig:sys}. To provide a comprehensive study, Fig.~\ref{fig:armchair} showcases the conductance plot and its FFT for the armchair edge configuration. Fig.~\ref{fig:armchair}(a) presents the conductance plotted against the magnetic field for the same three configurations. Different from that for the systems with zigzag edges, the conductance is small at low magnetic fields and gradually increases to $1e^2/h$ at higher B. This phenomenon is related to the distinct band structures of the GNRs. Specifically, GNRs with zigzag edges are characterized as gapless due to the presence of a zero-energy edge state, while those with armchair edges possess a finite energy gap and our choice of Fermi energy is lower than this energy gap. As B increases, the lowest conductance band will become lower and flatter to form the Landau levels. Consequently, the conductance gradually increases as a function of B. In contrast to the case with the zigzag edge, the oscillation pattern displayed in the 3slits and 4slits extends to a wider range of magnetic fields, demonstrating the robustness of this phenomenon. The corresponding Fourier spectrum in Fig.~\ref{fig:armchair}(b) also reveals the presence of higher harmonics.

\begin{figure}[t!]
\centering    \includegraphics[width=0.48\textwidth]{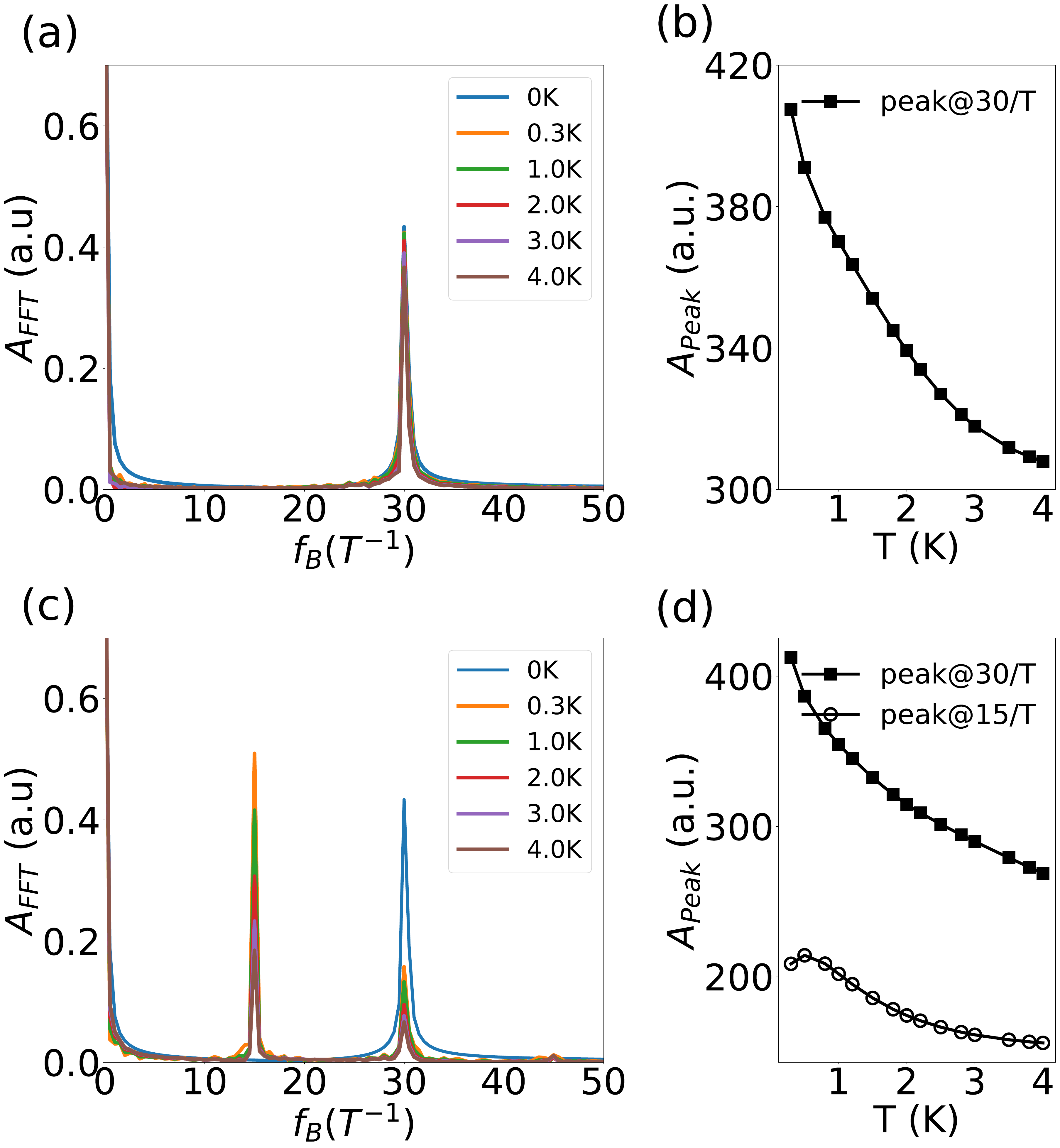}
\captionsetup{singlelinecheck=false, justification=justified}
\caption{\justifying 
(a) and (c) display the Fourier transforms for the 2slits and 3slits systems, respectively, calculated at finite temperatures ranging from 0.3 K to 4 K, with $E_F=0.1 meV$. The integrated amplitudes of the FFT peaks as functions of temperature are presented in (b) and (d) for the corresponding systems. Specifically, for the 2slits system, (b) displays the integrated amplitude of the peak at 30/T, with the integral in the frequency domain ranging from $(25-35)/T$. In contrast, for the 3slits system, (d) illustrates the integrated amplitudes at 15/T and 30/T, with the integral ranges of $(10-20)/T$ and $(25-35)/T$, respectively.}
\label{fig:cohlength}
\end{figure}

\section{Influence of temperature}
Extending our analysis to finite temperatures is essential to gain a comprehensive understanding of the multi-path quantum transport dynamics under realistic conditions. This investigation is particularly pertinent to practical applications of graphene-based devices, which often operate at nonzero temperatures.

To incorporate finite temperature effects, we need to integrate the product of the transmission and broadening function over an energy window ($E_F\pm 15k_BT$), with $k_BT$ representing the thermal energy~\cite{datta_1995}:
\begin{equation}
    G(B,E_F,T)=\frac{e^2}{h} \int T(E,B) F_{T}(E-E_F) dE
\end{equation}
where $F_{T}(E-E_F)= -\frac{\partial {f_{E_F}(E)}}{\partial{E}}$ denotes the thermal broadening function and $f_{E_F}(E)=[1+e^{(E-E_F)/k_BT}]^{-1}$ is the Fermi–Dirac distribution function. In standard cryogenic experiments, a  temperature of 4K corresponds to a thermal energy of approximately 0.3meV, providing a useful reference for understanding the energy scales involved in low-temperature simulations. In the calculation, we fix the Fermi energy at 0.1meV, and vary the temperature across the range from 0K to 4K. As shown in Fig.~\ref{fig:cohlength}(a,c), the integrated amplitude of the FFT peaks at 30/T reduces for both the 2slits and 3slits configurations. In contrast, the FFT peaks at 15/T do not decrease monotonously for the 3slits configuration. The peaks at 15/T may be enhanced at 0.3K. This nontrivial temperature effect is related to the interference of electron matter waves among nearest neighbor paths in the 3slits configuration. Additionally, the peak values at 15/T (without integration) are usually larger than those at 30/T, despite being narrower.

\section{Conclusion}
In summary, we have systematically studied the quantum transport dynamics of electrons within a multi-path AB interferometer constructed from parallel graphene nanoribbons. We unveil intriguing oscillatory behavior in conductance at low magnetic field strengths, reminiscent of the diffraction grating effect in optics, underscoring the interplay between electron trajectories, magnetic flux, and the quantum Hall effect. With increasing magnetic fields, certain pathways are blocked, and the system evolves into a two-path interferometer due to the formation of Landau levels and the associated chiral edge states. One important generalization of our work is to include the disorders that are always present in the real material. Disorders can affect the coherence length, mean-free length, and, subsequently, the phenomena of multi-path interference. Furthermore, our exploration can be expanded to encompass spin-orbit coupling in the tight-binding model and applied to other 2D materials. Our findings hold promise for the advancement of interferometry and quantum sensing technologies. They are poised to stimulate further theoretical and experimental investigations of multi-path interference with electronic matter waves.

\begin{acknowledgments}
We would like to thank Fan Zhang and Yang-Zhi Chou for stimulating comments and discussions. This work is supported by the ACC-New Jersey under Contract No. W15QKN-18-D-0040 and the Stevens Startup Fund.
\end{acknowledgments}

\bibliography{references}

\end{document}